\documentclass[letter]{aa} 

\usepackage{txfonts}
\usepackage{natbib}
\bibpunct{(}{)}{;}{a}{}{,} 
\usepackage{graphicx}
\usepackage{txfonts}
\newcommand{\Lx}{\mbox{$L_{\rm X}$}}

\newcommand{\lsim}{\raisebox{-.4ex}{$\stackrel{<}{\scriptstyle \sim}$}}
\newcommand{\msim}{\raisebox{-.4ex}{$\stackrel{>}{\scriptstyle \sim}$}}

\def\change{}
\def\nchange{}
\newcommand{\kq}{KQ~Vel}

\begin{document}

\title{{\em Chandra}'s X-ray study confirms that the magnetic standard \\ 
Ap star KQ~Vel hosts 
a neutron star companion\thanks{
The scientific results reported in this article are based 
on observations made by the {\em Chandra} X-ray Observatory
(ObsID 17745).}}
\titlerunning{X-ray observations confirm that KQ~Vel is an Ap+NS binary}
\authorrunning{Oskinova, Ignace, Leto, Postnov}

\author{
     Lidia M. Oskinova\inst{1,2}
\and Richard Ignace\inst{3}
\and Paolo Leto\inst{4}
\and Konstantin A. Postnov\inst{5,2}
}

\institute{
Institute for Physics and Astronomy, University Potsdam, D-14476 Potsdam,
Germany
\and
Department of Astronomy, Kazan Federal University, Kremlevskaya Str 18, Kazan, Russia
\and
Department of Physics \& Astronomy, East Tennessee
State University, Johnson City, TN, 37614, USA
\and
NAF - Osservatorio Astrofisico di Catania, Via S. Sofia 78, I-95123
Catania, Italy
\and
Sternberg Astronomical Institute, M.V. Lomonosov Moscow
University, Universitetskij pr. 13, 119234 Moscow, Russia
}

\date{Received <date> / Accepted <date>}

\abstract
{
KQ~Vel is a peculiar A0p star with a strong surface magnetic field of
about 7.5~kG.  It has a slow rotational period of nearly 8 years.
\citet{2015A&A...575A.115B} detected a binary companion of uncertain 
nature, and suggested it could be a neutron star or a black hole.  
}
{
We analyze  X-ray data obtained by the {\em
Chandra} telescope to ascertain information about the stellar magnetic field
and/or interaction between the star and its companion.  
}
{\nchange
We confirm previous X-ray detections of \kq\ with a relatively
large X-ray luminosity of $2\times 10^{30}$\,erg\,s$^{-1}$. X-ray 
spectra suggest the presence of hot gas at $> 20$~MK and, possibly, 
of a non-thermal component. {
\change X-ray light curves are variable, but better quality data are 
needed to determine periodicity if any.}  
}
{
We interpret X-ray spectra as a combination of two
components: the non-thermal emission arising from the  
aurora on the A0p star and the hot thermal plasma filling the 
extended shell surrounding the ``propelling''  neutron star.  
}
{
We explore various alternatives, but a hybrid model involving the
stellar magnetosphere along with a hot shell around the propelling neutron
star seems most plausible.{\nchange We speculate that \kq\ was originally a triple system, 
and the Ap star is a merger product}. We conclude that \kq\ is an intermediate-mass
binary consisting of a strongly magnetic main sequence star and a neutron star. 
}


\maketitle

\section{Introduction}

KQ Vel (HD 94660, HR 4263) {\nchange is a nearby star at $d\approx 114$\,pc with} a long
history of study. {\nchange It was} first identified as a chemically peculiar A~star by
\cite{1959PASP...71...48J}, while strong  surface magnetic field was detected 
by \cite{1975PASP...87..961B}. {\nchange The current estimates} show that the
field strength is in excess of 7.5~kG \cite[e.g.,][]{2017A&A...601A..14M}.  
The star is an exceedingly slow rotator, with a rotation period of 
about 2800\,d (Table\,1). 

\citet{2015A&A...575A.115B} determined that the star has a complex magnetic 
field, strongly non-solar abundances, with large overabundance of Fe-peak and 
rare-earth elements, and shows remarkable radial velocity variations with a 
period $\sim 840$\,d. The binary companion is not seen in optical, which led 
\citet{2015A&A...575A.115B} to suggest the first detection of a  
compact companion with mass $\msim \, 2\,M_\odot$ for a main sequence magnetic star. 


Among all A-type stars so far detected  in X-rays, 
KQ~Vel is the second most X-ray luminous \citep{2016AdSpR..58..727R}.  
The record holder 
is KW~Aur, where very soft X-ray emission likely arises from a white dwarf 
companion \citep{2007A&A...475..677S}.  
This begs the question of whether
the exceptional X-ray production from KQ~Vel also arises from 
its compact companion. 
In addition, a popular model for explaining 
X-ray emissions from
magnetic stars is the magnetically confined wind shock (MCWS) model of
\cite{1997A&A...323..121B}.   {\nchange This model has been developed 
to explain X-ray emissions from the magnetic A0P star IQ~Aur,  however
it fails to explain why some magnetic Ap stars are X-ray dim   \citep{2016AdSpR..58..727R}}.  
The MCWS model requires the presence of radiatively  driven 
stellar winds. Very little is known about the winds of non-supergiant
A stars \citep{Babel1996, Krticka2019}. What is clear, however, that the winds 
of these stars are very weak. 

Recently, radio studies of strongly magnetic chemically
peculiar Bp stars revealed variable polarimetric
behavior and radio continua consistent with non-thermal processes resulting
from auroral mechanisms
\citep[e.g.,][]{2017MNRAS.467.2820L,2018MNRAS.476..562L,2020MNRAS.493.4657L}.
\citet{2018A&A...619A..33R} suggested that the auroral mechanisms operates
also in strongly magnetic Ap stars, such as  CU~Vir, which is detected in
X-rays with $\Lx\approx3\times 10^{28}$\,erg\,s$^{-1}$ and has hard 
X-ray emissions with $T_{\rm X}\approx 25$\,K.


We undertook a study of the X-ray emission from KQ~Vel measured by the {\em
Chandra} X-ray telescope to clarify its origin:
could the emission arise from (a) the compact companion, (b) the
star's magnetosphere, or (c) an interaction between the weak but magnetized wind
of KQ~Vel and its compact companion.  In section~\ref{sec:obs} we
describe the new X-ray data.  A discussion of the results is given
in section~\ref{sec:disc} toward resolving the origin of the X-rays.
Section~\ref{sec:conc} presents concluding remarks, while detailed model 
calculations are presented in Appendix.

\section{X-ray properties of \kq}
\label{sec:obs}


\begin{table}
\caption{Stellar Properties of \kq$^{*}$}
\label{tab:star}
\begin{tabular}{lc}
\hline \hline
Sp.\,Type  & A0p EuSiCr \\
Distance & 114 pc \\
Temperature  & 11\,300~K \\
Radius & 2.53 $R_\odot$ \\
Luminosity, $L_{\rm bol}$  & $3.5\times10^{35}$ erg\,s$^{-1}$  \\
Mass & $3.0 \pm 0.2~M_\odot$ \\
Magnetic field strength & 7500~G\\
Rotation period, $P_{\rm rot}$  & 2800 d\\
Orbital period, $P_{\rm orb}$ & 840 d\\ 
{\nchange Eccentricity, $e$} & {\nchange 0.36} \\
\hline\hline  
\end{tabular}\\
\small{* from \citet{2015A&A...575A.115B}  and references therein}
\end{table}


%
%

\subsection{X-ray spectra}	\label{sub:spectra}

\kq\ was observed by the ACIS-I instrument on board the {\em Chandra}
X-ray telescope on 2016-08-20  for 25~ks (ObsID 17745).  We retrieved
and analyzed these archival X-ray data using the most recent
calibration files.  The spectrum and the light curve were extracted 
using standard procedures from a region with diameter $\approx 7\arcsec$.  
The background area was chosen in a nearby area free of X-ray sources. 
{\nchange The net count rate is $0.1$\,s$^{-1}$. The pile-up is $\approx 13$\%\ and 
does not significantly affect the spectral fitting results. 
Throughout the text, the X-ray properties of \kq\ are reported in 
the 0.3--11.0\,keV band}. 

\begin{figure}
\resizebox{\hsize}{!}{\includegraphics[angle=-90,origin=c]{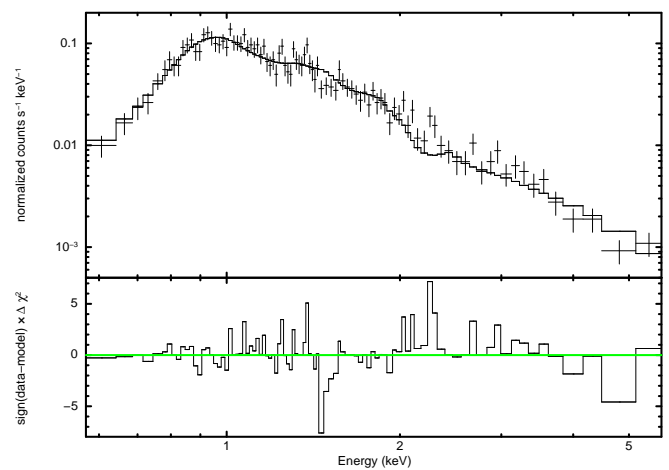}}
\caption{{\nchange A part of the {\em Chandra}  ACIS-I spectrum of \kq\ 
in the 0.5--9 \,keV energy range and  with error bars corresponding to 3$\sigma$ is 
shown by data points. The solid line shows the best fit 2T 
({\it apec}) plus power-law model. 
The model was fit over the 0.3--11 \,keV energy band.} 
The model parameters are given in Table~\ref{tab:par_xray}.  }
\label{fig:2T}
\end{figure}

To analyze the spectra we used the standard X-ray spectral fitting
software {\sc xspec} \citep{arnaud1996}. The abundances were scaled 
relative to solar values according to \citet{Asplund2009}. {\nchange 
\kq\ is a chemically peculiar star. For example, \cite{2015A&A...575A.115B} 
find overabundances among the iron group elements by factors of $1000$.
However allowing for non-solar abundances during spectral fitting does not
improve the fits of these low-spectral resolution data, and for now we adopt 
solar abundances.

Spectral fits of similar statistical quality   were obtained using 
two different spectral models: (1) purely thermal plasma model with 
a two-temperature (2T) collisional ionization equilibrium ({\em apec}) 
components, with the hottest plasma at $\approx
30$~MK; (2) combined thermal and non-thermal model which assumed a 
thermal, {\em apec}, component plus a power-law  (see Fig.\,\ref{fig:2T}). 
In this spectral model, the thermal plasma component has a 
temperature of $\approx 10$~MK. 
The parameters of best fit models  are listed in Table~\ref{tab:par_xray}.

For both models, when the newest calibration files that account for the 
contamination on the ACIS-I detector are used, the neutral hydrogen column 
density, $N_{\rm H}$, is consistent with being negligible. 
Based on \cite{1970A&A.....4..234F} and \cite{2001ApJ...558..309D}, the
intrinsic color\footnote{We used tables found at
www.stsci.edu/$\sim$inr/intrins.html that are based on the work of
the cited authors.} of an A0~star is $(B-V)_0 = -0.08$.  The observed
$(B-V)_{\rm obs} = 0.02$, formally implying a negative reddening
$E(B-V) = (B-V)_{\rm obs} - (B-V)_0$.  We interpret this as a
very low level of interstellar reddening, and, correspondingly, a
low neutral H~column density in the direction of \kq; this is consistent
with the results from X-ray spectral modeling.}
\begin{table}
\begin{center}
\caption{\nchange X-ray Spectral Model Fitting}
\label{tab:par_xray}
\footnotesize
\begin{tabular}{ll}
\hline
\hline
\multicolumn{2}{l}{Thermal model ({\em tbabs(apec+apec)})}\\
\hline
$kT_1$ & $0.81\pm 0.04$ keV  \\
$EM_1$ & $(3.5\pm 0.8)\times 10^{52}$ cm$^{-3}$ \\
$kT_2$ & $2.5\pm 0.2$ keV  \\
$EM_2$ & $(7.1\pm 0.4)\times 10^{52}$ cm$^{-3}$   \\ 
$\langle k T \rangle \equiv \sum_i k T_i \cdot EM_i / \sum_i EM_i$ & 1.9 keV   \\
reduced $\chi^2$ for 95 d.o.f. & 1.1 \\
Flux$^a$ & $1.3\times 10^{-12}$ erg cm$^{-2}$ s$^{-1}$   \\
\hline

\multicolumn{2}{l}{Thermal plus a power-law  model, 
({\em tbabs(apec+power)})}\\
\hline
$kT$ & $0.88\pm 0.04$ keV   \\
$EM$ & $(3\pm 0.5)\times 10^{52}$ cm$^{-3}$  \\
$\alpha$ & $2.5\pm0.2$   \\
$K$ (at 1 keV) & $(2.4\pm 0.3)\times10^{-4}$ keV$^{-1}$ cm$^{-2}$ s$^{-1}$    \\
reduced $\chi^2$ for 95 d.o.f. & 1.1 \\
Flux$^a$ & $1.5 \times 10^{-12}$ erg cm$^{-2}$ s$^{-1}$   \\
\hline
$L_{\mathrm X}^{\mathrm b}$ & $3 \times 10^{30}$ erg s$^{-1}$  \\
$\log L_{\mathrm X} / L_{\mathrm {bol}}$ & $-5$   \\
\hline

\end{tabular}
\begin{list}{}{}
\item[$^{\mathrm{a}}$] observed; in the 0.3--11 keV band
\end{list}
\end{center}
\end{table}


From the Rankine-Hugoniot condition for a strong shock, the post-shock
temperature is given by $T \approx 14~{\rm MK}\times (\varv_{\mathrm w} /
10^3~{\mathrm {km}}~{\mathrm s}^{-1})^2$, for $v_{\rm w}$ the wind
speed.  Achieving a temperature of 30\,MK  would require 
the speeds exceeding $1500$\,km s$^{-1}$.  This is significantly 
larger than  expected for A0V stars; e.g.\ the empirically 
estimated terminal wind velocities of the main sequence B-type stars 
do not exceed 1000\,km\,s$^{-1}$ \citep{Prinja1989}.
We favor the combined, thermal and non-thermal plasma model as
better motivated physically. By analogy with the Bp stars, the 
non-thermal X-rays in the spectrum of \kq\ 
could be explained as bremsstrahlung emission from a non-thermal
electron population which are also responsible
for the gyro-synchrotron stellar radio emission. When they impact
the stellar surface,  X-rays are radiated by thick-target bremsstrahlung
emission.  This physical process is well understood; in particular,
the spectral index $\alpha$, of the non-thermal photons, can be
related to the spectral index $\delta$, of the non-thermal electron
population, by the simple relation $\delta=\alpha +1$ \citep{brown_71}.
Then, using the best fit X-ray spectrum of \kq, the spectral index
of the non-thermal electron energy distribution is  $\delta=3.5$ .


\begin{figure}
\resizebox{\hsize}{!}{\includegraphics[origin=c]{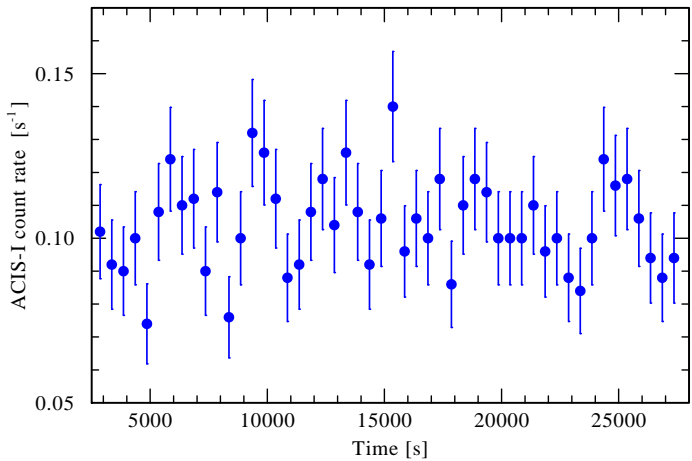}}
\caption{ X-ray light curve of \kq. 
Data are for the 0.3--10.0\,keV
(1.24--62\,\AA) energy band, where the background was subtracted.
The horizontal axis denotes the time after the beginning of the
observation in seconds. The data were binned to 500\,s.  The vertical
axis shows the count rate as measured by the ACIS-I camera. The
error bars (1$\sigma$) correspond to the combination of the error
in the source counts and the background counts.}
\label{fig:xlc}
\end{figure}

An X-ray light curve of \kq\ shown in Figure~\ref{fig:xlc} is variable.  
The power-density spectrum was calculated  using 
a Fourier algorithm \citep{Horne1986}. 
The largest peak in power density appears at a period of $P_{\rm X}=3125\pm 500$\,s, 
but with a  false-alarm probability of 50\%.  {\change Consequently,
the  period is not statistically significant. However, the periodicities at amplitudes 
too low to be detected with the current data cannot be ruled out.}




\section{Discussion}
\label{sec:disc}

The {\em Chandra} data reveal that KQ~Vel is extraordinary  
in its X-ray properties -- not only this star is among  most luminous, it is 
also the hardest X-ray source among Ap type stars. 
In the following sections, we explore several
different models in an attempt to explain the observed X-ray
properties of the \kq:  the overall level of X-ray emission,
and its hardness and  variability.

\subsection{Auroral X-ray emission of a magnetic Ap star}
Recent improvements of MCWS models account for the auroral 
emission, and are capable of explaining jointly radio and X-ray 
observations of magnetic B2Vp stars \citep{2017MNRAS.467.2820L,2020MNRAS.493.4657L}. 

Auroral model was also successfully applied to the Ap star CU\,Vir 
\citep{2018A&A...619A..33R}.
However, there are principal differences between  CU\,Vir  and \kq. 
The former is a fast  rotator ($P_{\rm rot}\approx 0.52$\,d) whereas 
the later rotates very slowly ($P_{\rm rot} \approx 2800$\,d). Furthermore, 
the polar field of CU\,Vir  is about two times smaller than for KQ\,Vel \citep{2014A&A...565A..83K}.
A high field strength is a key parameter for accelerating plasma electrons or
protons up to relativistic energies.  Upon impinging on the stellar
surface at the polar caps, this population of non-thermal particles
(electrons/protons) radiate non-thermal X-rays by thick target
bremsstrahlung. The viewing geometry of KQ\,Vel is favorable to see 
always the same stellar magnetic pole \citep{2015A&A...575A.115B}.
On the other hand, stellar rotation has a key role in the generation of the
non-thermal electrons.
In slow rotators, such as \kq, the centrifugal effect that helps the 
trapped  material to break the magnetic field lines becomes negligible. 
Therefore, the thermal electrons at the equatorial current sheet
have much lower density compared to the case of fast rotation. Indeed, all ApBp 
stars with detected auroral radio or/and X-ray emission are fast rotators. 
In the case of the B2V star $\rho$\,Oph\,A, which has its X-ray emission
 mainly sustained by auroral mechanism, {\nchange the ratio
between the X-ray and the radio luminosity is  
$L_{\rm X}/L_{\nu,\, \rm radio}=10^{14}$\,Hz \citep{2020MNRAS.493.4657L}.  
Adopting this ratio, the expected radio luminosity of \kq\ is at the mJy
level. 
}
 
 

We believe that the auroral mechanism can be responsible for 
the non-thermal spectral component in the X-ray spectrum of \kq. 
However, to explain its thermal X-ray emission, additional mechanisms 
must be invoked. 


\subsection{Binarity with a non-degenerate companion}

KQ~Vel is a binary star.  The analysis of radial velocity variations by
\cite{2015A&A...575A.115B} suggests a companion of around $2~M_\odot$.
They argue that such a companion could not be an optical star,
or spectral features betraying its nature would
have been detected.  They conclude that the companion is  a compact
object -- either a neutron star (NS) or a black hole (BH).  Besides, 
they consider a possibility of a hierarchical system, perhaps with two
lower mass companions in a tighter binary.

In this latter scenario, the unusually hard and luminous X-rays could 
be coronal in nature, arising either or both of the companions.  However, 
the KQ Vel X-ray luminosity is 3--5 times greater than the star II Peg 
\citep{2004ApJ...609L..79T} or $\sigma$~Gem \citep{2013ApJ...768..135H}, 
which are considered strong coronal sources (D.\,Huenemoerder, private comm).

{\nchange The companion could be an active RS~CVn type binary 
consisting of coronal stars. These binaries  have 
typical X-ray luminosities of $10^{30}-10^{31}$~erg s$^{-1}$  
\citep{1981ApJ...245..671W, Dempsey1993}, i.e.\ similar to the observed in \kq. 
\citet{Montes1995} studied the behavior of activity indicators, such as H$\alpha$, 
Ca\,II\,K, and X-ray emission in a sample of 51 chromospherically active binary 
systems. It was demonstrated that the activity indicators are correlated. Assuming
that X-ray luminosity of \kq\ is due to a hidden RS~CVn-type companion, one would 
expect it contribution to H$\alpha$, H$\epsilon$, and Ca\,II\,K lines observed in 
the \kq\ spectra. However, \citet{2015A&A...575A.115B} do not report peculiarities 
in these lines, especially in \ Ca\,II\,K (see their section\,6.5). 
Hence, for now, we discard a RS~CVn companion as a possible explanation 
for the X-ray emission from KQ~Vel's.  }

\citet{Han2003} performed binary population synthesis  that 
predicted a large population of A0+sdB binary stars. {\nchange The typical 
mass of an sdOB star is lower than the companion mass in \kq, moreover 
the X-ray luminosity of \kq\ is significantly higher than that of  
sdOB  stars \citep{Mereghetti2016}}. At present, we consider this 
scenario unlikely. 


%

\subsection{Binary with a degenerate companion}

\citet{2015A&A...575A.115B} proposed that \kq\ has a NS or a BH companion.  
An immediate question is  whether this could explain the remarkable X-ray emission of \kq.

\subsubsection{\nchange Accreting neutron star, middle age pulsar, or a cataclysmic variable?}
\label{sec:accr}

The wind mass-loss rate from an A0 star is very small  \citep{Babel1996}. 
An upper limit of $10^{-12}~M_\odot$~yr$^{-1}$ 
has been determined for CU~Vir \citep{Krticka2019}, which
has the same spectral type as \kq.  From spindown
considerations, using equation~(25) from \cite{2009MNRAS.392.1022U},
\kq\ would achieve its slow rotation after $\sim 200$~Myr at a mass-loss
rate of $10^{-13}~M_\odot$~yr$^{-1}$, while it would take $\sim 2$~Gyr
at $10^{-15}~M_\odot$~yr$^{-1}$. \citet{Kochukhov2006} estimated the age 
of \kq\ as 260\,Myr (assuming a single star evolution). If the secondary 
which we observe now as \kq, was re-rejuvenated due to binary mass exchange {\nchange or  merger}
, it may be even younger. Hence we roughly assume 
$\dot{M}\sim 10^{-13}~M_\odot$~yr$^{-1}$ as consistent with both the 
stellar age and spectral type of \kq. 
 
{\nchange Assuming a NS accreting donor's
stellar wind  \citep{Davidson1973}}, the resulting X-ray luminosity of \kq\ is 
$\sim 10^{27}$\,erg\,s$^{-1}$, i.e.\ a few orders of magnitude below the observed. 
Furthermore, the strong magnetic field of \kq\ dominates over its 
feeble stellar wind. As a result, the plasma-$\beta$ is very low \citep{alt1969}. 
Using equation (5) from \citet{Oskinova2011} and adopting as an upper limit on the 
wind speed 1000\,km\,s$^{-1}$, we roughly estimate the {\em Alfv\'en} 
radius of \kq\ as $> 150\,R_\ast$ -- well within the orbital separation
of $245\,R_\ast$. It means that only a small fraction of the wind ($\lsim 1\%$)
which consists of neutral hydrogen and metals can escape the stellar 
magnetosphere and feed the NS.  Hence, we can rule out direct accretion 
onto a compact object as an explanation for the observed X-ray luminosity.

%


{\nchange Some middle-age rotation powered pulsars can emit X-rays at the level 
we observe in \kq\ \citep{Becker1997, Kargaltsev2005}. 
{\change In this case, the pulsed X-ray emission modulated with a NS spin period of 
less than a dozen seconds is expected. Unfortunately, our {\em Chandra}'s data  
are not well suited to search for such pulsation. 
The X-ray spectra of \kq\ are consistent  with the presence of thermal optically 
thin plasma. This is quite different from the X-ray spectra of rotation powered pulsars.} Hence, 
we believe that the current data do  not support the presence of a middle aged pulsar in the \kq\ system. 

Among commonly detected Galactic X-ray sources are cataclysmic variables (CV) 
\citep[e.g.][]{Revnivtsev2006}. These interacting binaries consist of an accreting 
white dwarf (WD) and a low-mass donor star filling its Roche lobe. CVs
could have X-ray luminosities and hard X-ray spectra comparable with 
those observed in \kq.  Many CVs have relatively weak and broad emission 
lines in their spectra, which, combined with the CVs being much fainter than A stars, 
could potentially lead to an undetectable Ap+CV triple system\footnote{This possibility 
was suggested by the anonymous reviewer}. 

{\change This, however, meets difficulties given the vast difference in ages 
among  A-type stars and typical CVs. The age of optical star in \kq\ is 
$< 300$\,Myr. At this age a WD may already be formed but generally  CVs belong to 
much older stellar populations. The Ap star in the \kq\ system could be a result of a 
merger, and hence being rejuvenated (see sect.\,\ref{sec:evol}). 
Nevertheless, even in this case, it would require a fine tuning to produce an AOV+CV system.  
Another consideration against a CV nature of the \kq\ companion, is that some CVs are associated 
with novae. For such a bright (V=6.11\,mag) and 
nearby star, a historical nova would have a good chance to be noticed if happen at 
the time of existing southern hemisphere records. However, since the typical 
recurrence time for classical novae is hundreds of years, a nova outburst associated with \kq\ 
might have been missed.  Therefore, while we currently do not favor the CV nature of the \kq\ companion, 
this explanation cannot be ruled out.}
\subsubsection{Propelling neutron star}

When a magnetized NS is embedded in stellar wind with very low mass-loss rate, 
such as in \kq, the quasi-spherical subsonic regime of accretion sets in 
\citep{Shakura2012}. 
The wind is gravitationally captured from a volume much 
larger than determined by the outer boundary of 
the corotating portion of the NS magnetosphere, $R_{\rm A}$.  
Denote the characteristic radius of the volume from which material 
can be captured as $R_{\rm B}$.  
In a rotating NS, the Keplerian radius is defined as 
$R_{\rm K} = (GM_{\rm NS}/\omega^2)^{1/3}$
with $\omega=2\pi/P_{\rm NS}$ for the NS spin period.
If the condition $R_{\rm K}\le R_{\rm A}$ is met, the accretion
is throttled through the ``propeller effect'' \citep{Illarionov1975}. 
When the captured 
gas lacks the specific angular momentum to pass into the corotating 
magnetosphere, the {\nchange accumulated} wind material can form an extended, 
quasi-spherical envelope of hot, X-ray emitting gas.  

{\nchange
In this situation, the NS is not accreting, and the captured stellar 
wind remains in the NS gravitational potential. 
The energy source preventing the gas from cooling is 
the mechanical power supplied by a propelling NS, which is mediated by the magnetic forces 
and convection in the shell. In this sense, the mass-loss from the optical star is not the 
parameter directly regulating the shell structure. Much more important (see Appendix)
is the stellar wind velocity which determines the outer shell radius. The density at 
the shell base, $\rho_{\rm A}$,  is related 
to the total mass of the gas in the shell which determines the observed emission 
measure ($EM$) (using a crude analogy, the gas density near the ground on Earth 
is determined by the total mass and temperature of the Earth atmosphere).
Therefore, there is no direct relation of $\rho_{\rm A}$ with the poorly known 
stellar wind mass-loss rate.  X-ray emission from the gas shell is sustained by 
the NS spin-down and not by wind accretion, hence the  orbital eccentricity is irrelevant. 
}

It is possible to obtain a self-consistent solution for the hot envelope
using the parameters of \kq. The detailed model calculations are described
in the Appendix. We adopt the NS mass as $M_{NS} = 1.5 M_\odot$, and
use constraints from the X-ray observations: 
$\Lx \approx 2\times 10^{30}$\,erg\,s$^{-1}$, $EM \approx 5 \times 10^{52}$\,cm$^{-3}$, 
$\bar{T} \approx 1$\,keV (Table 2). Then, following the theory of quasi-spherical 
accretion, 
the derived  inner and outer radii ($R_{\rm A}$ and $R_{\rm B}$), 
the mass of the shell $M_{\rm sh}$,
the wind speed of the A0p star, and the
magnetic moment ($\mu$) of the NS are: 
$\varv_{\rm w} \approx 500$ km s$^{-1}$, 
$R_{\rm A} \approx 0.1 R_\odot$, $R_{\rm B} \approx 2.2 R_\odot$,  $\mu_{30} \approx 3$, 
$M_{\rm sh}\approx 1 \times 10^{-14}\,M_\odot$, $\mu_{30} \approx 3$,
where $\mu_{30}$ is the magnetic moment of the NS normalized
to $10^{30}$\,G\,cm$^3$ -- a typical value for a NS.  

The derived quantities are in good agreement with the 
expected properties of the \kq\ system.  The wind speed, 500\,km\,s$^{-1}$, 
is appropriately realistic.  The derived $R_{\rm A}=0.1 R_\odot$
implies if the NS has spin period $P_{\rm NS}\lsim 260$\,s, it
will be in a propeller state. Such spin period is quite  
reasonable for a NS that has not spun-down to the accretion stage, given
the weak stellar wind of the A0p companion and a large binary separation.

While the accretion power is far too small to explain the observed X-ray 
luminosity of \kq, the propelling NS injects significant amount of 
energy via interaction of the rotating magnetosphere and the matter in the 
shell. The maximum power which is provided by a propelling NS is 
$\approx 2\pi(\mu^2/R_{\rm A}^3)/P_{\rm NS}$ \citep{Shakura2012}. Inserting 
parameters derived for \kq,  the NS should have  
$P_{\rm NS}\lsim 100$\,s, to explain the observed X-ray luminosity.  
According to the Eqs.\,(\ref{eq:sd}) and (\ref{eq:sd100}), the 
NS spin-down can power the hot shell surrounding NS in the \kq\ system 
at the present level  for at least $10^5$~years.

Furthermore, the theory of quasi-spherical accretion naturally explains
the variability seen in the X-ray light curve (Figs.\,\ref{fig:xlc}). The time 
scale of observed X-ray 
variability,  $P_{\rm X} \sim 3000$\,s, is much shorter than either 
the orbital period of the binary ($P_{\rm orb}=840$\,d) or the spin 
period of the A0p star ($P_{\rm rot}=2800$\,d).  At the same time, 
it is at least an order of magnitude longer than the predicted NS 
spin period, $P_{\rm NS}\lsim 100$\,s. However, 
according to the Eq.\,(\ref{eq:tff}), the free-fall time a hot 
convective shell surrounding the NS is 
$t_{\rm ff}\simeq 3200~{\rm s}\, (\varv_{\rm w}/500~{\rm km\,s}^{-1})^{-3}$,
i.e.\ very similar to the observed variability time scale in \kq.  


Finally, one should consider the special properties of \kq, specifically  
the strong magnetic field of the donor star and hence its magnetized 
wind. The estimates show (see the Appendix) that the magnetic reconnection 
of the accretion flow and the NS field provides an important source of 
plasma heating maintaining the hot envelope around the NS in the \kq\ binary 
system.    

\subsection{\nchange Evolutionary considerations:  was \kq\ a triple? }
\label{sec:evol}
{\nchange
The \kq\ system has a wide and eccentric orbit. If a NS is present
in the system, a supernova (SN) must have occurred without disrupting
the binary. In the absence of an additional kick velocity, the
orbital eccentricity following the SN event would become $e=\Delta
M/(M_{\rm NS}+M_{\rm opt})$. For the parameters of KQ~Vel
(Tab.~\ref{tab:star}), $\Delta M=0.36\cdot(3+1.5)=1.6\,M_\odot$.
Hence, the NS progenitor mass can be roughly estimated as $1.6+1.5\approx
3\,M_\odot$. This likely was a He-star, stripped during the
mass-exchange event. The initial mass of the NS progenitor can then
be estimated \citep[see][]{Postnov2014} as $M_{\rm He}=0.1 M_{\rm
init}^{1.4}$, or a total mass of $\sim 11\,M_\odot$. Taking into
account the long orbital period, the progenitor system might have
been a C-type (i.e., a wide binary). In such cases the mass transfer
can be highly unstable and likely non-conservative, and the hydrogen
envelope will be lost without adding mass to the $3\, M_\odot$
secondary companion.  A sort of common envelope (CE) may have
occurred, but the efficiency of CE in wide binaries is uncertain.

This alternative scenario may help to shed light on the origin of
\kq\ system and its peculiar properties, particularly the strongly
magnetic nature of the optical star. It has long been believed, and
recently confirmed by numerical modeling, that strongly magnetic
massive stars result from stellar mergers \citep{Schneider2019}.
In this case, \kq\ was originally a triple system, where the
present-day Ap star is the merger product, possibly of a W~UMa type
system.  The NS is the tertiary remnant. If the initially less
massive tertiary gained mass during the Roche lobe overflow by
either of the primary binary components, the orbital separation
would have increased.

The eccentric Kozai-Lidov mechanism \citep{Naoz2014} operating in
triple systems causes strong inclination and eccentricity fluctuations,
and leads to the tightening of the inner binary. The merger product
of the inner binary is observed today as the magnetic Ap star.  The
tertiary, with a mass in the range $8-10\,M_\odot$ may have
evolved to an electron-capture SN, without disrupting the system, and
formed the NS we observe today.  }

{\change The proximity of \kq\ to Earth implies that the \kq-like systems cannot be 
exotic and exceptionally rare. The life-time of a propelling NS is $\sim 10^5-10^6$\,yr, 
i.e.\ significantly less than a few\,$\times 10^8$\,yr lifetime of an A0-type star. After 
the NS has spun down and the propeller stage has ended, the NS will enter the 
accretion regime, however given a very small accretion rate, its X-ray 
luminosity will be small (sect.\,\ref{sec:accr}). Given a large orbital separation, 
the binarity may be easily missed during a routine spectroscopic analysis. Future 
work on astrometric catalogs, e.g.\ {\em Gaia},  will be undoubtedly useful to shed more light on the 
population of wide intermediate mass binaries with X-ray dim NS companions. 
}

\section{Conclusions} 
\label{sec:conc}

{\em Chandra} X-ray observations of the strongly magnetic binary star 
\kq\ revealed that its properties are exceptional for an Ap-type star:  
high X-ray luminosity, hard  spectrum, and 
variable X-ray light-curve. 
The observed spectrum can be well fit either by a multi-temperature 
thermal plasma spectral model, or by  a hybrid model
consisting of thermal  and a non-thermal emission components. We prefer the 
former spectral model, and explain the non-thermal component as being 
due to the auroral mechanism, i.e.\ similar to the Ap star CU\,Vir. 

Exploring several different interpretations, we concluded that the 
theory of quasi-spherical accretion onto propelling  NS with 
$P_{\rm NS} \lsim 100$\,s and $\mu_{30}\approx 3$ best 
describes 
the hot plasma temperature, its emission  measure, 
and the  time scale of X-ray variability observed in \kq.
Thus, new X-ray {\em Chandra} observations strongly support the 
\citet{2015A&A...575A.115B} suggestion on the presence of a compact companion 
in \kq. We conclude that the compact companion is a NS in the propeller regime. 

\kq\ is the first known strongly magnetic Ap\,+\,NS binary. Confirmation of the 
existence of such objects has important consequences for our understanding 
of binary evolution and accretion physics. 

\begin{acknowledgements}
{\nchange Authors are grateful to the anonymous referee for a very useful 
report which strongly improved the paper, and for the suggestions 
for future work on this interesting system.}
Authors thank Dr.\,H. Todt for sharing the statistical models. 
LMO  acknowledges financial support by the Deutsches Zentrum f\"ur
Luft und Raumfahrt (DLR) grant FKZ 50 OR 1809, and partial support
by the Russian Government Program of Competitive Growth of Kazan
Federal University.  The work of KAP is partially supported ny RFBR
grant 19-02-00790.
\end{acknowledgements}


\begin{thebibliography}{45}
\expandafter\ifx\csname natexlab\endcsname\relax\def\natexlab#1{#1}\fi

\bibitem[{{Altschuler} \& {Newkirk}(1969)}]{alt1969}
{Altschuler}, M.~D. \& {Newkirk}, G. 1969, \solphys, 9, 131

\bibitem[{{Arnaud}(1996)}]{arnaud1996}
{Arnaud}, K.~A. 1996, in Astronomical Society of the Pacific Conference Series,
  Vol. 101, Astronomical Data Analysis Software and Systems V, ed.
  {G.~H.~Jacoby \& J.~Barnes}, 17--+

\bibitem[{{Asplund} {et~al.}(2009){Asplund}, {Grevesse}, {Sauval}, \&
  {Scott}}]{Asplund2009}
{Asplund}, M., {Grevesse}, N., {Sauval}, A.~J., \& {Scott}, P. 2009, \araa, 47,
  481

\bibitem[{{Babel}(1996)}]{Babel1996}
{Babel}, J. 1996, \aap, 309, 867

\bibitem[{{Babel} \& {Montmerle}(1997)}]{1997A&A...323..121B}
{Babel}, J. \& {Montmerle}, T. 1997, \aap, 323, 121

\bibitem[{{Bailey} {et~al.}(2015){Bailey}, {Grunhut}, \&
  {Landstreet}}]{2015A&A...575A.115B}
{Bailey}, J.~D., {Grunhut}, J., \& {Landstreet}, J.~D. 2015, \aap, 575, A115

\bibitem[{{Becker} \& {Truemper}(1997)}]{Becker1997}
{Becker}, W. \& {Truemper}, J. 1997, \aap, 326, 682

\bibitem[{{Borra} \& {Landstreet}(1975)}]{1975PASP...87..961B}
{Borra}, E.~F. \& {Landstreet}, J.~D. 1975, \pasp, 87, 961

\bibitem[{{Brown}(1971)}]{brown_71}
{Brown}, J.~C. 1971, \solphys, 18, 489

\bibitem[{{Cowie} {et~al.}(1981){Cowie}, {McKee}, \&
  {Ostriker}}]{1981ApJ...247..908C}
{Cowie}, L.~L., {McKee}, C.~F., \& {Ostriker}, J.~P. 1981, \apj, 247, 908

\bibitem[{{Davidson} \& {Ostriker}(1973)}]{Davidson1973}
{Davidson}, K. \& {Ostriker}, J.~P. 1973, \apj, 179, 585

\bibitem[{{Dempsey} {et~al.}(1993){Dempsey}, {Linsky}, {Fleming}, \&
  {Schmitt}}]{Dempsey1993}
{Dempsey}, R.~C., {Linsky}, J.~L., {Fleming}, T.~A., \& {Schmitt}, J.~H.~M.~M.
  1993, \apjs, 86, 599

\bibitem[{{Ducati} {et~al.}(2001){Ducati}, {Bevilacqua}, {Rembold}, \&
  {Ribeiro}}]{2001ApJ...558..309D}
{Ducati}, J.~R., {Bevilacqua}, C.~M., {Rembold}, S. r.~B., \& {Ribeiro}, D.
  2001, \apj, 558, 309

\bibitem[{{Fitzgerald}(1970)}]{1970A&A.....4..234F}
{Fitzgerald}, M.~P. 1970, \aap, 4, 234

\bibitem[{{Han} {et~al.}(2003){Han}, {Podsiadlowski}, {Maxted}, \&
  {Marsh}}]{Han2003}
{Han}, Z., {Podsiadlowski}, P., {Maxted}, P.~F.~L., \& {Marsh}, T.~R. 2003,
  \mnras, 341, 669

\bibitem[{{Horne} \& {Baliunas}(1986)}]{Horne1986}
{Horne}, J.~H. \& {Baliunas}, S.~L. 1986, \apj, 302, 757

\bibitem[{{Huenemoerder} {et~al.}(2013){Huenemoerder}, {Phillips}, {Sylwester},
  \& {Sylwester}}]{2013ApJ...768..135H}
{Huenemoerder}, D.~P., {Phillips}, K. J.~H., {Sylwester}, J., \& {Sylwester},
  B. 2013, \apj, 768, 135

\bibitem[{{Illarionov} \& {Sunyaev}(1975)}]{Illarionov1975}
{Illarionov}, A.~F. \& {Sunyaev}, R.~A. 1975, \aap, 39, 185

\bibitem[{{Jaschek} \& {Jaschek}(1959)}]{1959PASP...71...48J}
{Jaschek}, M. \& {Jaschek}, C. 1959, \pasp, 71, 48

\bibitem[{{Kargaltsev} {et~al.}(2005){Kargaltsev}, {Pavlov}, {Zavlin}, \&
  {Romani}}]{Kargaltsev2005}
{Kargaltsev}, O.~Y., {Pavlov}, G.~G., {Zavlin}, V.~E., \& {Romani}, R.~W. 2005,
  \apj, 625, 307

\bibitem[{{Kochukhov} \& {Bagnulo}(2006)}]{Kochukhov2006}
{Kochukhov}, O. \& {Bagnulo}, S. 2006, \aap, 450, 763

\bibitem[{{Kochukhov} {et~al.}(2014){Kochukhov}, {L{\"u}ftinger}, {Neiner},
  {Alecian}, \& {MiMeS Collaboration}}]{2014A&A...565A..83K}
{Kochukhov}, O., {L{\"u}ftinger}, T., {Neiner}, C., {Alecian}, E., \& {MiMeS
  Collaboration}. 2014, \aap, 565, A83

\bibitem[{{Krti{\v{c}}ka} {et~al.}(2019){Krti{\v{c}}ka}, {Mikul{\'a}{\v{s}}ek},
  {Henry}, {Jan{\'\i}k}, {Kochukhov}, {Pigulski}, {Leto}, {Trigilio},
  {Krti{\v{c}}kov{\'a}}, {L{\"u}ftinger}, {Prv{\'a}k}, \&
  {Tich{\'y}}}]{Krticka2019}
{Krti{\v{c}}ka}, J., {Mikul{\'a}{\v{s}}ek}, Z., {Henry}, G.~W., {et~al.} 2019,
  \aap, 625, A34

\bibitem[{{Leto} {et~al.}(2020){Leto}, {Trigilio}, {Leone}, {Pillitteri},
  {Buemi}, {Fossati}, {Cavallaro}, {Oskinova}, {Ignace}, {Krti{\v{c}}ka},
  {Umana}, {Catanzaro}, {Ingallinera}, {Bufano}, {Agliozzo}, {Phillips},
  {Cerrigone}, {Riggi}, {Loru}, {Munari}, {Gangi}, {Giarrusso}, \&
  {Robrade}}]{2020MNRAS.493.4657L}
{Leto}, P., {Trigilio}, C., {Leone}, F., {et~al.} 2020, \mnras, 493, 4657

\bibitem[{{Leto} {et~al.}(2017){Leto}, {Trigilio}, {Oskinova}, {Ignace},
  {Buemi}, {Umana}, {Ingallinera}, {Todt}, \& {Leone}}]{2017MNRAS.467.2820L}
{Leto}, P., {Trigilio}, C., {Oskinova}, L., {et~al.} 2017, \mnras, 467, 2820

\bibitem[{{Leto} {et~al.}(2018){Leto}, {Trigilio}, {Oskinova}, {Ignace},
  {Buemi}, {Umana}, {Ingallinera}, {Leone}, {Phillips}, {Agliozzo}, {Todt}, \&
  {Cerrigone}}]{2018MNRAS.476..562L}
{Leto}, P., {Trigilio}, C., {Oskinova}, L.~M., {et~al.} 2018, \mnras, 476, 562

\bibitem[{{Mathys}(2017)}]{2017A&A...601A..14M}
{Mathys}, G. 2017, \aap, 601, A14

\bibitem[{{Mereghetti} \& {La Palombara}(2016)}]{Mereghetti2016}
{Mereghetti}, S. \& {La Palombara}, N. 2016, Advances in Space Research, 58,
  809

\bibitem[{{Montes} {et~al.}(1995){Montes}, {Fernandez-Figueroa}, {de Castro},
  \& {Cornide}}]{Montes1995}
{Montes}, D., {Fernandez-Figueroa}, M.~J., {de Castro}, E., \& {Cornide}, M.
  1995, \aap, 294, 165

\bibitem[{{Naoz} \& {Fabrycky}(2014)}]{Naoz2014}
{Naoz}, S. \& {Fabrycky}, D.~C. 2014, \apj, 793, 137

\bibitem[{{Oskinova} {et~al.}(2011){Oskinova}, {Todt}, {Ignace}, {Brown},
  {Cassinelli}, \& {Hamann}}]{Oskinova2011}
{Oskinova}, L.~M., {Todt}, H., {Ignace}, R., {et~al.} 2011, \mnras, 416, 1456

\bibitem[{{Postnov} {et~al.}(2017){Postnov}, {Oskinova}, \&
  {Torrej{\'o}n}}]{2017MNRAS.465L.119P}
{Postnov}, K., {Oskinova}, L., \& {Torrej{\'o}n}, J.~M. 2017, \mnras, 465, L119

\bibitem[{{Postnov} \& {Yungelson}(2014)}]{Postnov2014}
{Postnov}, K.~A. \& {Yungelson}, L.~R. 2014, Living Reviews in Relativity, 17,
  3

\bibitem[{{Prinja}(1989)}]{Prinja1989}
{Prinja}, R.~K. 1989, \mnras, 241, 721

\bibitem[{{Raymond} {et~al.}(1976){Raymond}, {Cox}, \&
  {Smith}}]{1976ApJ...204..290R}
{Raymond}, J.~C., {Cox}, D.~P., \& {Smith}, B.~W. 1976, \apj, 204, 290

\bibitem[{{Revnivtsev} {et~al.}(2006){Revnivtsev}, {Sazonov}, {Gilfanov},
  {Churazov}, \& {Sunyaev}}]{Revnivtsev2006}
{Revnivtsev}, M., {Sazonov}, S., {Gilfanov}, M., {Churazov}, E., \& {Sunyaev},
  R. 2006, \aap, 452, 169

\bibitem[{{Robrade}(2016)}]{2016AdSpR..58..727R}
{Robrade}, J. 2016, Advances in Space Research, 58, 727

\bibitem[{{Robrade} {et~al.}(2018){Robrade}, {Oskinova}, {Schmitt}, {Leto}, \&
  {Trigilio}}]{2018A&A...619A..33R}
{Robrade}, J., {Oskinova}, L.~M., {Schmitt}, J.~H.~M.~M., {Leto}, P., \&
  {Trigilio}, C. 2018, \aap, 619, A33

\bibitem[{{Schneider} {et~al.}(2019){Schneider}, {Ohlmann}, {Podsiadlowski},
  {R{\"o}pke}, {Balbus}, {Pakmor}, \& {Springel}}]{Schneider2019}
{Schneider}, F. R.~N., {Ohlmann}, S.~T., {Podsiadlowski}, P., {et~al.} 2019,
  \nat, 574, 211

\bibitem[{{Schr{\"o}der} \& {Schmitt}(2007)}]{2007A&A...475..677S}
{Schr{\"o}der}, C. \& {Schmitt}, J.~H.~M.~M. 2007, \aap, 475, 677

\bibitem[{{Shakura} {et~al.}(2012){Shakura}, {Postnov}, {Kochetkova}, \&
  {Hjalmarsdotter}}]{Shakura2012}
{Shakura}, N., {Postnov}, K., {Kochetkova}, A., \& {Hjalmarsdotter}, L. 2012,
  \mnras, 420, 216

\bibitem[{{Syunyaev} \& {Shakura}(1977)}]{Syunyaev1977}
{Syunyaev}, R.~A. \& {Shakura}, N.~I. 1977, Soviet Astronomy Letters, 3, 138

\bibitem[{{Testa} {et~al.}(2004){Testa}, {Drake}, {Peres}, \&
  {DeLuca}}]{2004ApJ...609L..79T}
{Testa}, P., {Drake}, J.~J., {Peres}, G., \& {DeLuca}, E.~E. 2004, \apjl, 609,
  L79

\bibitem[{{Ud-Doula} {et~al.}(2009){Ud-Doula}, {Owocki}, \&
  {Townsend}}]{2009MNRAS.392.1022U}
{Ud-Doula}, A., {Owocki}, S.~P., \& {Townsend}, R. H.~D. 2009, \mnras, 392,
  1022

\bibitem[{{Walter} \& {Bowyer}(1981)}]{1981ApJ...245..671W}
{Walter}, F.~M. \& {Bowyer}, S. 1981, \apj, 245, 671

\end{thebibliography}

\appendix

\section{X-ray emission from a hot shell around a propelling 
neutron star in settling accretion regime}
\label{sec:app}

The X-ray emission properties from a hot shell around the magnetosphere
of a propelling neutron star (a model applied to the $\gamma$~Cas
phenomenon) were calculated in \cite{2017MNRAS.465L.119P} (see
Eqs.~(2)-(11) in that paper). The calculations were carried out for
the case of thermal bremsstrahlung cooling, which is valid for high
plasma temperatures $kT\gtrsim 4$~keV. At lower plasma temperatures,
the collisional cooling function becomes dominant, rapidly increasing
down to temperatures $\sim 0.01$~keV. To estimate the properties
of the shell in the temperature range $0.01 < kT<4$~keV, we can use
the analytical approximation for a fully ionized plasma with solar
abundances \citep{1976ApJ...204..290R,1981ApJ...247..908C}:

\begin{equation}
\Lambda = K_\mathrm{cool}\, T^{-0.6} ,
\end{equation}

\noindent with $K_\mathrm{cool}=6.2\times 10^{-19}$ in cgs units.
Below we shall replace the numerical power 0.6 with 3/5.

The total X-ray luminosity from an optically thin spherical shell
located between the magnetospheric radius $R_{\rm A}$ and the outer radius
$R_{\rm B}$ is

\begin{equation}
L_{\rm X}=\int\limits_{R_{\rm A}}^{R_{\rm B}}\,n_e^2 \,\Lambda\, 4\pi
\,r^2\,dr\,.
\label{eq:lx}
\end{equation}
\noindent Here we have assumed a completely ionized H gas, with
$n_{\rm H} = n_{\rm e} = \rho/m_{\rm H}$. For the outer boundary in
the integral (\ref{eq:lx}), we assume
the Bondi radius with $R_{\rm B}=2GM_{\rm NS}/\varv_{\rm w}^2$, where
$M_{\rm NS}$ is the NS mass, $v_{\rm w}$ is the velocity of the wind 
from the optical star, and neglecting the orbital motion of the NS.

The density and temperature profiles in quasi-stationary gas envelope
surround the NS magnetosphere \citep{Shakura2012} are:

\begin{eqnarray}
\rho(r) & = & \rho_{\rm A}\,\left(\frac{R_{\rm A}}{r}\right)^{3/2},~{\rm
and}\\
T(r) & = & T_{\rm A}\,\left(\frac{R_{\rm A}}{r}\right).
\end{eqnarray}

\noindent Using the virial temperature of a monoatomic gas at the
base of the shell $T_{\rm A}= GM_{\rm NS} m_{\rm H}/5k R_{\rm A}$, we obtain

\begin{equation}
\label{e:Lx}
L_{\rm X}=4\pi\frac{5}{3}K_\mathrm{cool}\left(\frac{ \rho_{\rm A}}
{m_{\rm H}}
\right)^2R_{\rm A}^3\left(\frac{GM_{\rm NS} m_{\rm H}}{5kR_{\rm B}}
\right)^{-3/5}\left[1-\left(\frac{R_{\rm A}}{R_{\rm B}}\right)^{3/5}\right].
\end{equation}

An important difference of Eq.~(\ref{e:Lx}) from the analogous
Eq.~(6) in \cite{2017MNRAS.465L.119P} is that, due to inverse
power-law temperature dependence of $\Lambda$, the X-ray luminosity
is almost fully determined by the combination $\rho_{\rm A}^2R_{\rm A}^3$
and
the stellar wind velocity $\varv_{\rm w}$ (which sets the Bondi radius
via $GM_{\rm NS}/R_{\rm B}=1/2 \varv_{\rm w}^2$). The ratio $R_{\rm
A}/R_{\rm B}$
is typically $\ll 1$, which enables us to omit the second term in the square
brackets in Eq.~\ref{e:Lx}. The $EM$ in the shell becomes:

\begin{eqnarray}
\label{e:EM}
&{EM}=\int\limits_{R_{\rm A}}^{R_{\rm B}}n_{\rm e}^2 4\pi r^2dr
=4\pi \left(\frac{ \rho_{\rm A}}{m_{\rm H}}\right)^2
R_{\rm A}^3\ln(R_{\rm B}/R_{\rm A})\nonumber\\
&=\frac{3}{5K_\mathrm{cool}}L_{\rm X}\left(
\frac{G\,M_{\rm NS}\, m_{\rm H}}{5k\,R_{\rm B}}\right)^{3/5}
\ln(R_{\rm B}/R_{\rm A})\,.
\end{eqnarray}

\noindent Noting that $GM_{\rm NS}/R_{\rm B}= \varv_{\rm w}^2/2$ and
substituting
the characteristic values $M_{\rm NS}=1.5 M_\odot$, $\varv_{\rm w}=10^8~{\rm
cm\,s^{-1}}\times \varv_8$, $L_{\rm X}=10^{30}~{\rm erg\,s^{-1}} \times
L_{30}$,
we arrive at

\begin{equation}
\label{e:EMn}
{EM}\approx 1.15\times 10^{52}\,L_{30}\,\varv_8^{6/5\,}
\ln(R_{\rm B}/R_{\rm A})~{\rm cm}^{-3}.
\end{equation}

\noindent The logarithmic factor here is usually on the order of
3--4. The $EM$  is observationally constrained, hence it is possible
to determine the mass contained in the shell, with

\begin{equation}
M_{\rm sh}=m_{\rm H}\,\int_{R_{\rm A}}^{R_{\rm B}}\,4\pi\,r^2\,n_{\rm H}(r)\,dr
= \frac{8\pi}{3}\,m_{\rm H}\,n_{\rm A}\,R_{\rm A}^3\,
(R_{\rm B}/R_{\rm A})^{3/2}.
\end{equation}
Using Eq.~\ref{e:Lx}, and the density and temperature profiles in the shell,
it is straightforward to derive the expression for the magnetospheric
radius $R_{\rm A}$:

\begin{eqnarray}
&R_{\rm A}=\left(\frac{20\,K_\mathrm{cool}}{m_{\rm H}^2}\right)
^{1/7} \left(
\frac{5K_2}{2\pi}\right)^{2/7}\left(\frac{\mu^4}{L_{\rm X}}\right)^{1/7}
\left(\frac{G\,M_{\rm NS} \,m_{\rm H}}{5k\,R_{\rm
B}}\right)^{-3/35}\nonumber \\
&\simeq 4\times 10^9\,\mu_{30}^{4/7}L_{30}^{-1/7}\varv_8^{-6/35}~{\rm
cm}.
\label{eq:ra}
\end{eqnarray}

\noindent Here $\mu=10^{30}~{\rm G\,cm^3}\,\mu_{30}$ is the NS
magnetic moment, $K_2\approx 7.7$ is a numerical factor that takes
into account the difference of the NS magnetosphere from a pure
dipole shape (c.f., \cite{Shakura2012}). For typical wind velocities,
$R_{\rm B}\sim 4\times 10^{10}\,v_8^{-2}$~cm, and therefore our assumption
$R_{\rm B}/R_{\rm A}\gg 1$ is justified.

The temperature at the base of the shell is
$T_{\rm A} \propto R_{\rm A}^{-1}\simeq
10~{\rm keV}\,\mu_{30}^{-4/7}\,L_{30}^{1/7}\,\varv_8^{6/35}$. However,
relevant to the fit of the observed spectrum is neither the maximum
nor the minimum temperature, but the average temperature as weighted
by $EM$. The predicted average temperature, $\bar{T}$, is given
by

\begin{equation}
\bar{T} = \frac{\int_{T_{\rm A}}^{T_{\rm B}}\,T(r)\,dEM}{\int_{T_{\rm
A}}^{T_{\rm B}}\,dEM}.
\end{equation}

\noindent With $dEM = n^2(r)\,4\pi\,r^2\,dr$, and using the preceding
expressions, along with $R_{\rm B} \gg R_{\rm A}$, the weighted average
temperature becomes

\begin{equation}
\bar{T} \approx \frac{T_{\rm A}}{\ln (R_{\rm B}/R_{\rm A})}.
\label{eq:avt}
\end{equation}

\noindent Given that the denominator is typically of order a few, the
characteristic thermal temperature expected from the hot shell model is a
factor of a few smaller than the temperature at the inner boundary of
the shell.

Related to spectral fitting is the column density distribution of
material in the shell. Even if hydrogen is completely ionized in the hot
shell, it is standard to express the column as $N_{\rm H}$. Much of the
absorption at the higher energies derives from photoabsorption by
metals.
Along a radial through the annual shell and assuming $R_B/R_A\gg 1$, 
the column density 
is:

\begin{equation}
N_{\rm H} = \int_{R_{\rm A}}^{R_{\rm B}}\,n_{\rm H}(r)\,dr 
\simeq 2 R_{\rm A}\rho_{\rm A}/m_{\rm H}  
\label{eq:nh}
\end{equation}



Equations (\ref{e:Lx}, \ref{e:EM}, \ref{eq:avt}) enable us to express three 
independent model parameters, $\rho_{\rm A}$, $R_{\rm A}$ and $R_{\rm B}$ 
(or $\varv_{\rm w}$) through  
the values which are directly derived from observations: \Lx, $EM$, and 
$\bar{T}$. Then, Eq.\,\ref{eq:ra} can be used to estimate the NS magnetic field.  

Finally, let us check whether the shell can remain hot under assumed form of
the cooling function that increases along the radius as $\Lambda\sim
T^{-0.6}\sim r^{0.6}$.  The free-fall time of the hot envelope is given by
\begin{equation}
t_{\rm ff} = \frac{R_{\rm B}^{3/2}}{\sqrt{2\,G\,M_{\rm NS}}} 
= \frac{2\,G\,M_{\rm NS}}{\varv_{\rm w}^3}\simeq 3200~{\rm s}\,
(\varv_{\rm w}/500~{\rm km\,s}^{-1})^{-3}.
\label{eq:tff}
\end{equation}
The plasma cooling time in the isentropic shell can
be expressed as
\begin{eqnarray}
&t_{\rm cool}(r)=\frac{3kT}{n_{\rm e}\Lambda}\simeq 10^3 
[\mathrm{s}]\frac{T_{\rm A}^{8/5}(R_{\rm A}/r)^{8/5}}{n_{\rm a}(R_{\rm A}/r)^{3/2}} 
\nonumber\\
&=10^3
[\mathrm{s}]\left(\frac{R_{\rm A}}{r}\right)^{1/10}
\left[\frac{\ln(R_{\rm B}/R_{\rm A})}{3}\right]^{1/2}
\left(\frac{\mathrm{EM}}{5\times
10^{52}\,\mathrm{cm}^{-3}}\right)^{-1/2}
\end{eqnarray}
which is almost constant across the shell. However,
this time is dangerously close to the free-fall time
$t_{\rm ff}\sim 3\times 10^3[\mathrm{s}](r/R_{\rm B})^{3/2}$. This suggests that if
there were no additional plasma heating in the shell, the captured gas would
rapidly cool below $R_{\rm B}$ to form a cold dense 'dead' disk around the
magnetosphere
\citep{Syunyaev1977}, and no hot convective shell would be formed.

Luckily, in the case of magnetized wind the magnetic reconnection can heat
up plasma. Indeed, the reconnection time in a magnetized plasma blob of size
$l$, mass $m_{\rm b}\sim \rho_{\rm b} l^3$ and magnetic field $B_{\rm b}$ 
can be written as
$t_{\rm r}\sim l/v_{\rm r}$, where $v_{\rm r}$ is the reconnection
rate scaling as the
Alfv\'en velocity $\varv_{\rm A}\sim B_{\rm b}/\sqrt{\rho_{\rm b}}$. 
Therefore, $t_{\rm r}\sim l\sqrt{\rho_{\rm b}}/B_{\rm b}$, and using 
the magnetic flux conservation $B_{\rm b}^2=const$
we arrive at 
$t_{\rm r}\sim l^3\sqrt{\rho_{\rm b}}\sim m_{\rm b}/\sqrt{\rho_{\rm b}}$. The
magnetic reconnection heating is effective if $t_{\rm r}/t_{\rm cool}\sim
(m_{\rm b}\rho)/T_{\rm b}^{8/5} < 1$. By neglecting mass decrease of the falling 
blob (due to, for example, Kelvin-Helmholtz stripping), assuming the 
adiabatic blob
evolution (i.e. $\rho_{\rm b} T_{\rm b} l^{5/3}=const$) in the surrounding 
plasma with
pressure $P_{\rm e}\sim n_{\rm e}T_{\rm e}$ and pressure balance 
$P_{\rm b}\sim \rho_{\rm b} T_{\rm b}=P_{\rm e}$, we
arrive at $t_{\rm r}/t_{\rm cool}\propto r^\frac{189}{20}$ (here the adiabatic
scaling for the surrounding plasma density and temperature, Eqs. (A.3) and
(A.4), were applied). This means that the magnetic reconnection in freely
falling magnetized plasma blobs
can rapidly occur providing additional heat sustaining the hot convective
shell.
On top of the plasma heating, the magnetic blob reconnection results in the
generation of a $\sim 10\%$ non-thermal tail. Realistically, not all
free-falling blobs are magnetized, and part of them can cool down thus
increasing $N_{\rm H}$ relative to the hot plasma estimate
Eq. (\ref{eq:nh}) above.

The spin-down timescale of the propelling NS is
\begin{equation}
t_{\rm sd}=\omega_{\rm NS}/\dot\omega_{\rm NS}=
(I\omega^2_{\rm NS}/L_{\rm X})(1+\epsilon), 
\label{eq:sd}
\end{equation}
where $I$ is the NS moment of inertia, and $\epsilon=L_{{\rm mr}}/L_{{\rm sd}}$ 
is the ratio of the NS spin-down power, $I\omega_{\rm NS}\dot\omega_{\rm NS}$, to
the power supplied by magnetic reconnection. Then

\begin{equation}
t_{\rm sd}\simeq 2\times 10^5[\mathrm{yrs}]
\left(\frac{P_{\rm NS}}{100\,\mathrm{s}}\right)^{-2}
\left(\frac{L_{\rm X}}{10^{30}\,\mathrm{erg\,s}^{-1}}\right)(1+\epsilon).
\label{eq:sd100}
\end{equation}

\end{document}